\documentclass[12pt]{article}
\usepackage{a4}

\usepackage{amssymb}
\renewcommand{\leq}{\leqslant}
\renewcommand{\geq}{\geqslant}

\usepackage{graphicx}
\newcommand{\dataplot}[1]{\includegraphics[angle=-90,width=\textwidth]{#1}}

\newcommand{\fracd}[2]{\frac{\displaystyle #1}{\displaystyle #2}}

\makeatletter
\def\sech{\mathop{\operator@font sech}\nolimits}
\makeatother

\begin{document}

\title{Gravitational binding in 4D dynamical triangulation}

\author{Bas V. de Bakker\thanks{email: bas@astro.uva.nl} \and Centre
  for High Energy Astrophysics\\ Kruislaan 403, 1098 SJ Amsterdam, the
  Netherlands.  \and \addtocounter{footnote}{1} \\Jan
  Smit\thanks{email: jsmit@phys.uva.nl}\\ Institute for Theoretical
  Physics, University of Amsterdam\\ Valckenierstraat 65, 1018 XE
  Amsterdam, the Netherlands.}

\date{August 19, 1996}

\maketitle

\begin{abstract}
  In the dynamical triangulation model of four dimensional euclidean
  quantum gravity we investigate gravitational binding.  Two scalar
  test particles (quenched approximation) have a positive binding
  energy, thereby showing that the model can represent gravitational
  attraction.
\end{abstract}

{%
\thispagestyle{myheadings}
\renewcommand{\thepage}%
{\vbox to 0pt{\hbox{AIAP-1996-043}\kern 6pt\hbox{ITFA-96-10}\vss}}

\clearpage
}

\section{Introduction}
Dynamical triangulation is a discrete approach to the 
path integral for euclidean quantum gravity in which the  
euclidean spacetimes are constructed by `glueing' together
geometric objects such as equilateral hypercubes or simplices. 
Hypercubes were used at first \cite{We82} whereas the modern 
formulation which developed independently out 2D gravity uses simplices.
Results of numerical simulations of the commonly used 4D model were first
presented in \cite{AmJu92,AgMi92a}. 
As a model of gravity dynamical triangulation ought to have a scaling regime 
where it corresponds to semiclassical Einstein gravity.  One wishes to recover
classical euclidean spacetimes, Newton's potential and the formation
of gravitationally bound states.

It is not clear that a purely euclidean formulation should be able to
contain semiclassical gravity, because of the well known divergence
related to the unboundedness of the euclidean version of the
Einstein-Hilbert action.  In a semiclassical evaluation of the
euclidean path integral this can be dealt with by deforming the
integration over the conformal mode into the imaginary direction of
the complex plane \cite{HaGiPe}. 

The dynamical triangulation formulation is completely regular from the
start.  The model has a phase transition as a function of a parameter
which is proportional to the inverse bare Newton constant: $\kappa_2
\propto G_0^{-1}$.  The transition separates a phase with crumpled
spacetimes and very high effective dimensionalities, from an
`elongated' phase with effectively two-dimensional spacetimes with
characteristics of a branched polymer \cite{BaSm95a,AmJu95a}.  Near
the transition the model appears to produce classical $S^4$-like
spacetimes, in an intermediate distance regime, and there is evidence
for scaling, suggesting continuum behaviour\footnote{We use the phrase
`continuum behaviour' rather than `continuum limit' to allow
for situations in which the lattice distance is very much shorter than 
any physical distance, but for which the continuum limit would imply 
unwanted features, such as noninteraction due to triviality (a typical example
is the 4D Ising model formulation of $\phi^4$ theory).}\cite{BaSm95a,Biaea96}.
In the elongated phase the scaling degenerates into a branched polymer
version \cite{AmJu95a}.

It is of great interest to find out if this scaling region can be
described by an effective action of the type 
\begin{equation}
  S_{\rm eff} = \int d^4 x\sqrt{g}\, \left(
     \frac{\Lambda}{8\pi G} 
   - \frac{R}{16\pi G} + \zeta R^2 + \cdots\right).
  \label{Seff}
\end{equation}
The branched polymer phase is presumably an expression of the
conformal mode instability of the $-R/G$-part in this effective action
at scales of order $\sqrt{G}$, while the crumpled phase may correspond
to negative $G$. 

Recent evidence suggests that the transition is of first order 
\cite{Biaea96,Ba96}, instead of second order as thought previously.  
The first order nature of the transition need 
not stand in the way of continuum behaviour
as the examples of gauge-Higgs models show.
These models have a first order phase transition also in the 
continuum treatment. For dynamical triangulation
we suggested previously \cite{BaSm95a} that continuum 
behavior may be automatic, as in two dimensions\footnote{We recall
also an analogy with $Z(n)$ gauge theories, 
which have a Coulomb phase for $n\geq 5$, with massless photons in 
a region of parameter space, without the need for
tuning to a critical point. Gauge-Higgs models of this type can have 
furthermore a first order Coulomb-Higgs transition \cite{Jaea86}. 
The models approach the U(1)-Higgs model as $n \to \infty$.
Because of triviality a continuum {\em limit} will be noninteracting, but the 
models may still be regarded (for finite sufficiently large $n$) 
as a nonperturbative formulation of the abelian Higgs model in the continuum.}.  
It is a priori possible that continuum analysis of the effective theory 
(\ref{Seff}) 
also leads to a first order transition between phases with 
positive\footnote{In 
dynamical triangulation the transition 
actually appears to occur at small positive $1/G$, 
since the spacetimes found in its neighbourhood 
are $S^4$-like, cf.\ sect.\ 5.} 
and negative $1/G$. 
%
The above interpretations are very speculative and more analytical as well
as numerical work has to be done to establish or reject it.

It would be very interesting if we could measure the attraction of two
sources at a fixed distance.  This has been 
pursued in the Regge
calculus formulation of simplicial quantum gravity \cite{HaWi94}.
Such measurements could then be compared with the
simple Newtonian law or with quantum corrections to this law, which
have for instance been calculated in \cite{Mo95} and \cite{HaLi95}.
In dynamical triangulation a computation of the gravitational
potential appears nontrivial, because it is difficult in a fluctuating
spacetime to keep two heavy test masses at a fixed distance.

However, the formation of bound states out of two test particles can
be computed in a way that is customary in lattice field theory and we
shall report on such computations in this paper.  Preliminary
results have appeared earlier in \cite{BaSm93} and \cite{BaSm95b}.
We see this work as an important ingredient for the interpretation
of 4D dynamical triangulation.

\section{Binding in the continuum}

Consider a free scalar field $\phi$ with bare mass $m_0$ in a quantum
gravity background. In continuum language, the euclidean action of
this system is a sum of a gravitational and a matter part
\begin{eqnarray}
  S & = & S[g] + S[g,\phi] , \\
  S[g] & = &
  \frac{1}{16\pi G_0} \int d^4x \, \sqrt{g} \left( 2 \Lambda_0 - 
  R \right) , \\
  S[g,\phi] & = &
  \int d^4x \, \sqrt{g} \left( \frac{1}{2} g^{\mu\nu} \partial_\mu
  \phi \partial_\nu \phi + \frac{1}{2} m_0^2 \phi^2 \right) ,
\end{eqnarray}
where $\Lambda_0$ is the bare cosmological constant, $R$ is the scalar
curvature and $G_0$ is the bare Newton constant.

We take $\phi$ as a test particle here, i.e.\ the back reaction of the
field $\phi$ on the metric is not taken into account. In lattice QCD
this approximation is often called the quenched approximation, or
valence quark approximation, because it neglects diagrams with
internal quark (in our case $\phi$) loops.  For not too light quark
masses it turns out to give good results (for a discussion see e.g.\ 
\cite{Sh94b}).  In dynamical triangulation the inclusion of a scalar
field is no problem in principle (and appears to have little influence
on the gravity sector of the theory) \cite{AmBuJuKr93}, but the
enlargement of parameter space by $m_0$ is computationally costly.  A
continuum calculation of the gravitational attraction of a scalar
field in the quenched approximation was carried out in \cite{Mo95}.

We will use the following notation for expectation values of an
observable $A$.  On a fixed background geometry we can average over
configurations of the matter field
\begin{equation}
  \langle A \rangle _\phi = \fracd{ \int \mathcal D \phi \, A \exp (
    - S[g,\phi] ) }{ \int \mathcal D \phi \, \exp ( - S[g,\phi] ) } ,
\end{equation}
and we can average over metrics
\begin{equation}
  \langle A \rangle _g = \fracd{ \int \mathcal D g \, A \exp ( -
    S[g] ) }{ \int \mathcal D g \, \exp ( - S[g] ) } ,
\end{equation}
The quenched expectation value is then
\begin{equation}
  \langle A \rangle = \langle \langle A \rangle_\phi \rangle_g .
\end{equation}

We next consider propagators. In a fixed geometry, the one particle
propagator, denoted by $G(x,y;g)$, is defined as
\begin{equation}
  G(x,y;g) = \langle \phi_x \phi_y \rangle_\phi ,
  \label{simprop}
\end{equation}
and the two-particle propagator is simply the square of the one 
particle propagator.
%
Letting the metric fluctuate, we take the average of the propagators
over the different metrics.  Because of reparametrisation invariance,
the average $\langle G(x,y;g) \rangle_g$ can only depend on whether
$x$ and $y$ coincide or not.  Therefore, we look at averages at fixed
geodesic distance $r$,
\begin{equation}
  G(r) = \Biggl\langle \fracd{\int d^4 x \sqrt{g} \, G(x,y;g) \,
  \delta(d(x,y)-r)}{\int d^4 x \sqrt{g} \,
  \delta(d(x,y)-r)}\Biggr\rangle_g ,
  \label{scalprop}
\end{equation}
where $d(x,y)$ is the minimal geodesic distance between $x$ and $y$.
By translation invariance, the resulting $G(r)$ is independent of
$y$.  Similarly, we can define the geometry average of the
two-particle propagator as
\begin{equation}
  G^{(2)}(r) = \Biggl\langle \fracd{\int d^4 x \sqrt{g} \, G(x,y;g)^2
  \, \delta(d(x,y)-r)}{\int d^4 x \sqrt{g} \, \delta(d(x,y)-r)}
  \Biggr\rangle_g .
  \label{twoprop2}
\end{equation}

In (\ref{scalprop}) we first averaged over the volume at distance $r$
from the point $y$ and then averaged over the metrics.  Alternatively
we can integrate $G(x,y;g)$ over the volume at distance $r$, average
over metrics, and then divide by the same with $G\rightarrow 1$,
\begin{equation}
  G(r) =  \fracd{
    \left\langle \int d^4 x \sqrt{g} \, G(x,y;g) \, \delta(d(x,y)-r) 
    \right\rangle_g
    }{
    \left\langle \int d^4 x \sqrt{g} \, \delta(d(x,y)-r) \right\rangle_g.
    }
  \label{scalprop2}
\end{equation}
A similar ambiguity arises with purely geometric correlators
\cite{BaSm95c}.  Our physical intuition tends to favour the form
(\ref{scalprop}), and we shall later use its analogue in dynamical
triangulation.  However, if there is no pair $x,y$ with $d(x,y) = r$
for a given metric $g$ the expression becomes mathematically ill
defined, whereas (\ref{scalprop2}) has no such problem.  For the pure
geometry correlators this does not happen in practise for reasonable
$r$ and the difference between (\ref{scalprop}) and (\ref{scalprop2})
appears to be small \cite{Bialas}.

For a massive particle, we expect the propagator (\ref{scalprop}) to
fall off exponentially as 
\begin{equation}
G(r) = Z  r^\alpha \exp (-mr),
  \label{gofd}
\end{equation}
with some power $\alpha$ and renormalised mass $m$, which in general
will not equal the bare mass $m_0$.  This expression neglects finite
size effects and should be modified when looking at distances
comparable to a typical length scale in the system.

The two-particle propagator will behave similarly as
\begin{equation}
  G^{(2)}(r)  =  Z^{(2)} r^\beta \exp(-M r) ,
  \label{g2ofd}
\end{equation}
where $M$ is the energy of the two-particle state.  If this energy
turns out to be less than two times the mass of a single particle, the
difference can be interpreted as a binding energy between the
particles.  This would show gravitational attraction between them.

\section{Propagating in constant curvature}

Since the spacetimes to be used in the binding energy computations
have topology $S^4$ and the average spacetime is expected to be
homogeneous, it will be useful to know the properties of scalar field
propagators on spaces of constant curvature.  To calculate such
propagators we have to solve the equation
\begin{equation}
  (\square - m^2)G = 0,
  \label{propeq}
\end{equation}
with the boundary condition $G \rightarrow (4\pi^2 r^2)^{-1}$ as the
geodesic distance $r\rightarrow 0$. We assume spherical symmetry,
i.e.\ $G$ depends only on $r$. We need to distinguish three cases:
positive, zero and negative curvature.  For a space with constant
positive curvature, i.e.\ a four-sphere, equation (\ref{propeq}) can
be written as
\begin{equation}
  ( \partial_x + 3 \cot x ) \partial_x G - m^2 r_0^2 G = 0 ,
\end{equation}
where $x = r/r_0$, with $r_0$ the curvature radius.  The substitution
$z = \cos^2(x/2)$, which was used in \cite{AnMo91}, turns this into a
hypergeometric equation,
\begin{equation}
  \left[ z (1-z) \partial_z^2 + (2-4z) \partial_z - m^2 r_0^2 \right]
    G(z) = 0.
\end{equation}
For numerical evaluation we found it easiest to use a series
representation
\begin{equation}
  G(z) = N \sum_{k=0}^{\infty} c_k z^k.
\end{equation}
Setting $c_0 = 1$, the differential equation fixes the $c_k$ as
\begin{equation}
  c_k = \frac{(k+2)(k-1) + m^2 r_0^2}{k(k+1)} c_{k-1} ,
\end{equation}
If we demand that the singularity at the origin goes like $(4\pi^2
r^2)^{-1}$, this fixes the normalisation to be
\begin{equation}
  N = \frac{m^2 r_0^2 - 2}{16\pi r_0^2 \cosh \left( \pi \sqrt{ m^2
  r_0^2 - \frac{9}{4}} \right) } .
\end{equation}
One can easily check that this function also conforms to the
differential equation at the point opposite the origin, where $x =
\pi$.  When the mass is so small that $m^2 r_0^2 < 9/4$, the above
formulas may be analytically continued such that
\begin{equation}
  N = \frac{m^2 r_0^2 - 2}{16\pi r_0^2 \cos \left( \pi
  \sqrt{\frac{9}{4} - m^2 r_0^2} \right) } .
\end{equation}

The negative curvature case gives similarly
\begin{equation}
  ( \partial_x + 3 \coth x ) \partial_x G - m^2 r_0^2 G = 0 .
\end{equation}
Using the substitution $z = \sech x$, we get
\begin{equation}
  \left[ (1-z^2) z^2 \partial_z^2 + (-2-2z^2) z \partial_z - m^2 r_0^2
  \right] G(z) = 0 .
\end{equation}
In this case we use the series
\begin{equation}
  G(z) = N \sum_{k=0}^{\infty} c_k z^{k+\mu},
\end{equation}
and the recurrence equation is now
\begin{equation}
  c_k = \frac{(k+\mu-2)(k+\mu-1)}{(k+\mu)(k+\mu-3) - m^2 r_0^2}\, 
  c_{k-2} .
\end{equation}
We can read off that we have to choose
\begin{equation}
  \mu = \frac{3}{2} + \sqrt{\frac{9}{4} + m^2 r_0^2} ,
\end{equation}
to avoid generating negative powers of $z$.  Setting again $c_0 = 1$
and demanding that $G$ goes like $(4\pi^2 r^2)^{-1}$ for small $r$
gives us the normalisation
\begin{equation}
  N = \frac{\Gamma(\mu/2) \Gamma(\mu/2 + 1/2)} {4\pi^2 r_0^2 \;
  \Gamma(\mu - 1/2)} .
\end{equation}

The flat case is the easiest.  Using the spherical symmetry results in
\begin{equation}
 \left( \partial_r + \frac{3}{r} \right) \partial_r G(r) - m^2 G(r) =
   0 ,
\end{equation}
which can be directly solved in terms of a Bessel function as
\begin{equation}
  G(r) = \frac{m K_1(mr)}{4\pi^2 r} ,
\end{equation}
using the condition that $G(r)$ goes like $(4\pi^2 r^2)^{-1}$ for
small $r$.

\begin{figure}[t]
  \dataplot{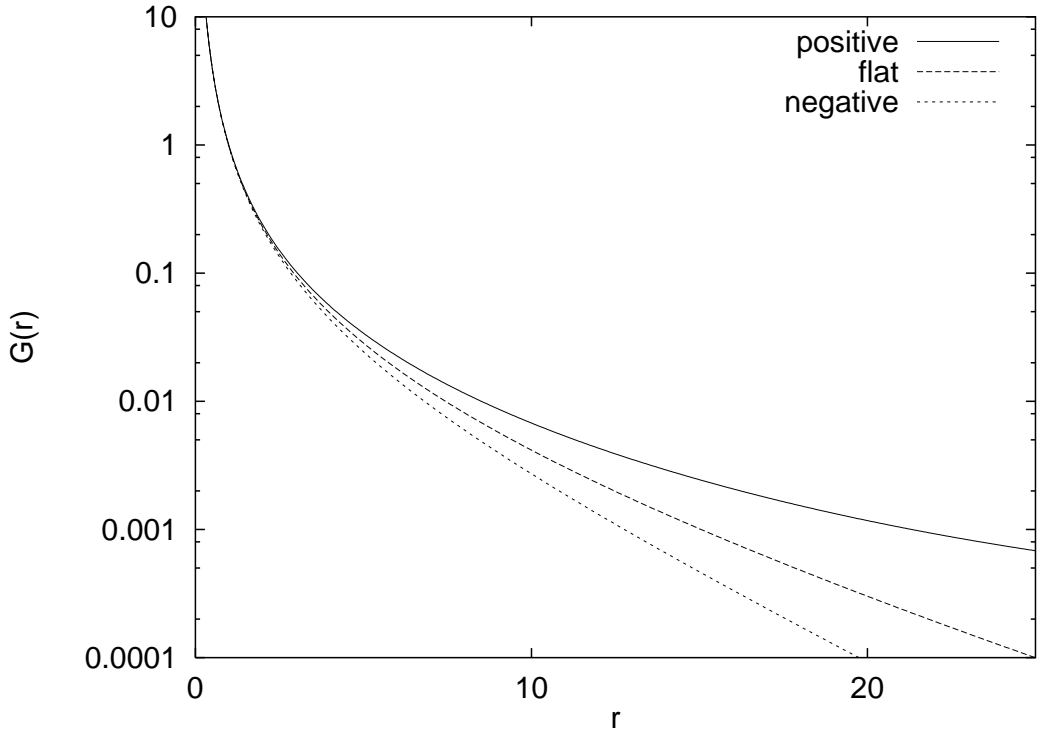}
  \caption{Propagators on spaces with constant curvature 
    $R=12/r_0^2$ (upper), $R=0$ (middle) and $R=-12/r_0^2$ (lower),
    with $r_0=13$ and $m=0.15$.}
  \label{curvature}
\end{figure}

We have plotted an example of the three cases in
figure~\ref{curvature}, using a curvature radius $r_0$ of $13$ and a
mass $m$ of $0.15$.  These parameters where chosen for later
comparison with dynamical triangulation results.

\section{Dynamical triangulation}

In the dynamical triangulation model of four dimensional euclidean
quantum gravity the path integral over metrics on a certain manifold
is defined by a weighted sum over all ways to glue four-simplices
together at the faces \cite{AmJu92,AgMi92a}.  This idea was first
formulated in \cite{We82}, using hypercubes instead of simplices.  In
four dimensions the analogue of the continuum gravitational action is
\begin{eqnarray}
  S[g] & = & \frac{1}{16 \pi G_0} \int d^4x \, \sqrt{g} \left( 2
  \Lambda_0 - R \right) \\ & \to & \kappa_4 N_4 - \kappa_2 N_2 ,
  \label{action}
\end{eqnarray}
where $N_2$ and $N_4$ are the number of triangles and four-simplices
respectively.

The partition function of the model for a fixed volume (fixed number
of four-simplices $N_4$) is given by
\begin{equation}
  Z(N,\kappa_2) = \sum_{{\cal T}(N_4 = N)} \exp(\kappa_2 N_2).
  \label{partfunc}
\end{equation}
The sum is over all ways to glue $N$ four-simplices together, such
that the resulting complex satisfies the manifold condition, with some
fixed topology which is usually (as well as in this article) taken to
be $S^4$.  The coupling constant $\kappa_2$ is proportional to the
inverse of the bare Newton constant:
\begin{equation}
  \kappa_2 = V_2 / 4G_0 ,
\end{equation}
where $V_2$ is the area of a triangle.

It turns out that the model has two phases 
\cite{AgMi92b,AmJuKr93,Br93,CaKoRe94a}.  For low $\kappa_2$ the system
is in a crumpled phase, where the average number of simplices around a
vertex is large and the average distance between two simplices is
small.  In this phase the volume within a distance $r$ appears to
increase exponentially with $r$, a behaviour like that of a space with
constant negative curvature.  At high $\kappa_2$ the system is in an
elongated phase and resembles a branched polymer.  As is the case with
a branched polymer, the (large scale) internal fractal dimension is 2.
The transition between the two phases occurs at a critical value
$\kappa_2^c$ which depends somewhat on $N$.  The phase transition
appears to be of first order \cite{Biaea96,Ba96}.  At the transition
the spaces behave on the average in several
respects like the four dimensional sphere \cite{BaSm95a}.  Some of the
evidence for this will be reviewed in the next section.

\section{Spacetimes near the transition}

We have performed numerical simulations of four dimensional dynamical
triangulation, according to the partition function (\ref{partfunc}).
We used systems of about $32000$ simplices and the topology of the
four-sphere.  To keep the number of simplices around the desired
value, we added a quadratic term to the action as was described in
\cite{AgMi92b,CaKoRe94a,BaSm94b}.

In ref.~\cite{BaSm95a} we studied the euclidean spacetimes generated
in numerical simulations of the model by measuring the number of
simplices $N'(r)$ at geodesic distance $r$.  (The geodesic distance
$d_{xy}$ between two simplices with centres $x$ and $y$ is defined as
the minimum number of links on the dual lattice between $x$ and $y$.)
We fitted $N'(r)$ in an intermediate distance regime by a form $c
\sin^{d-1} (r/r_0)$ corresponding to a $d$-sphere of radius $r_0$.
For $\kappa_2$ near the transition $\kappa_2^c(N)$ this gave $d
\approx 4$, which we took as evidence for classical $S^4$ behaviour at
the distance scale involved.  Below the transition (i.e.\ 
$\kappa_2<\kappa_2^c$) $d$ rises steeply to large values while above
the transition $d$ falls rapidly to the branched polymer value $d =
2$.

\begin{figure}[t]
  \dataplot{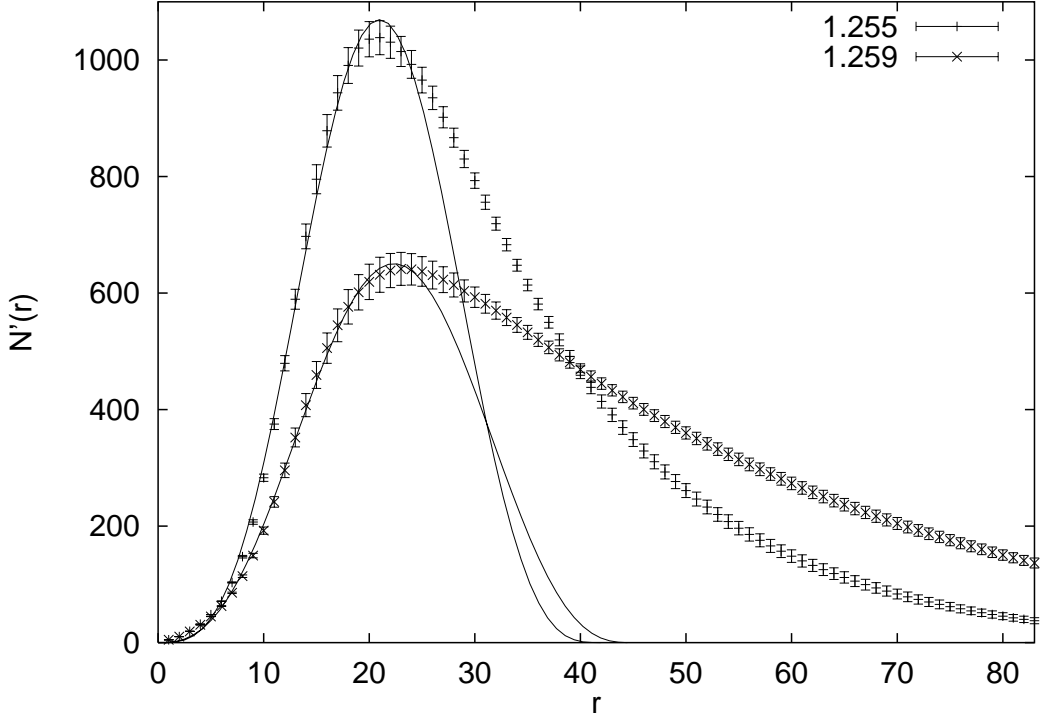}
  \caption{Fits to $N'(r)$ of the form $a\sin^{d-1}(r/r_0)$ near 
    the transition. Upper curve: $\kappa_2 = 1.255$, $d = 4.24(4)$,
    $r_0 = 13.35(7)$; lower curve: $\kappa_2 = 1.259$, $d = 3.67(7)$,
    $r_0 = 14.2(2)$.}
  \label{s4fig}
\end{figure}

To get a feeling for the geodesic distances which will appear later in
binding energy computations, we show in fig.~\ref{s4fig} the quantity
$N'(r)$ for a system of $N = 32000$ simplices at $\kappa_2 = 1.255$
and $1.259$.  The value $1.255$ is very near $\kappa_2^c(N)$ as
defined by the position of the maximum in the susceptibility
\begin{equation}
  \frac{\partial^2 \ln Z(N,\kappa_2)}{N\partial \kappa_2^2} \approx
  N \left[ \left\langle \frac{{N_2}^2}{{N_4}^2} \right\rangle -
  \left\langle \frac{N_2}{N_4} \right\rangle^2 \right] .
\end{equation}
The $\sin^{d-1} (r/r_0)$-fit gives $d=4.24(4)$ and is seen to be
reasonable in the region $6\leq r\leq 24$.  The value 1.259 is our
$\kappa_2$ closest to the phase transition for which the data fit the
four-sphere also reasonably well with $d = 3.67(7)$, in the region $5
\leq r \leq 25$.  The overall shape of $N'(r)$ is quite asymmetrical,
which is presumably due to branching fluctuations at larger distances.

In \cite{BaSm95a} we also introduced an effective curvature $R_{\rm
  eff}(r)$ to describe the curvature at scales much larger than the
lattice scale.  The argument $r$ of $R_{\rm eff}(r)$ is meant to
approach zero provided that, and as long as, $R_{\rm eff}(r)$ is
stationary.  This does not seem to happen in the elongated phase,
while near the transition and in the crumpled phase we found that
$R_{\rm eff}$ had indeed a stationary point (minimum).  For smaller
$r$-values $R_{\rm eff}$ rises steeply when $r$ gets smaller.  We
called this region of large $R_{\rm eff}$ the `planckian regime'.

\begin{figure}[t]
  \dataplot{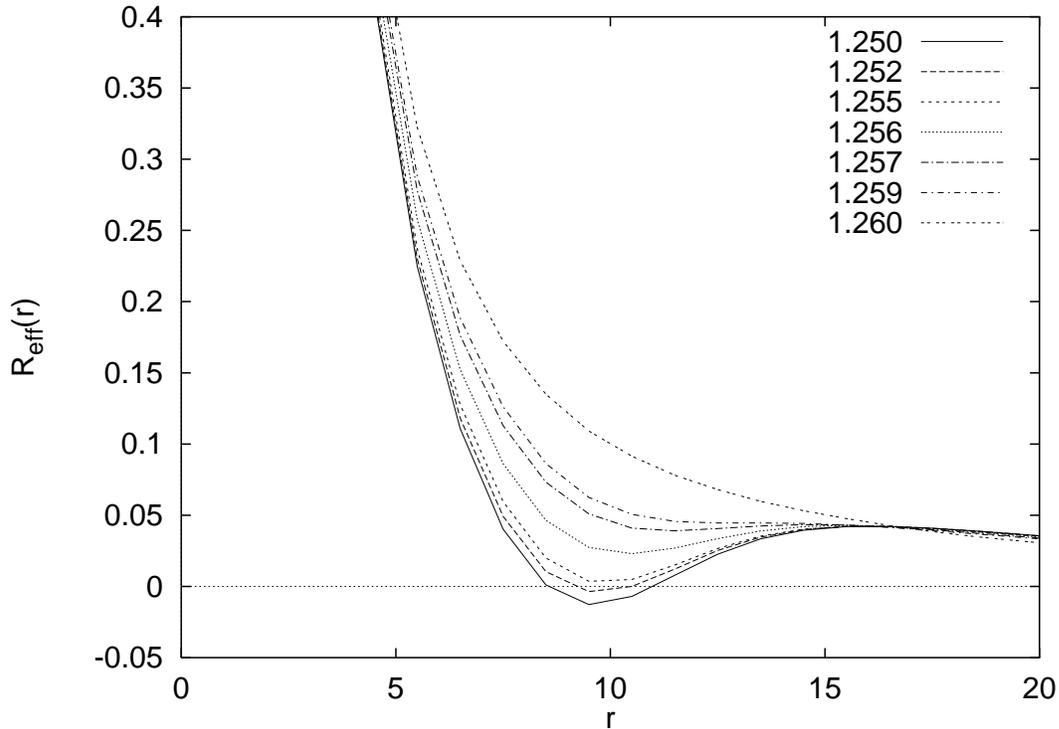}
  \caption{Effective curvature plots for $\kappa_2 = 1.250$--$1.260$.}
  \label{Refffig}
\end{figure}

Fig.~\ref{Refffig} shows $R_{\rm eff}$ for $N = 32000$ and several
values of $\kappa_2$. For $\kappa_2 < 1.255$ we see negative curvature
at the minimum of $R_{\rm eff}$.  A homogeneous space with constant
negative curvature is unbounded, so the maximum in $N'(r)$ is
evidently due to finite size effects.  We take the `planckian regime'
as the small $r$ region, roughly ending at the minimum of $R_{\rm
  eff}$.  For $\kappa_2 = 1.255$ the effective curvature is nearly
zero at the minimum, while positive for larger $\kappa_2$.  The
minimum has roughly turned into an inflection point at $\kappa_2 =
1.259$, and it has disappeared altogether for larger $\kappa_2$.

Fig.~\ref{s4fig} suggests positive curvature with a curvature radius
of $r_0 \approx 21/(\pi/2) \approx 13$.  This corresponds to some
average near the minima in fig.\ \ref{Refffig}.  The minima will show
somewhat smaller curvature: $R_{\rm eff}(r_{\rm min}) \approx 0$,
$0.045$, for $\kappa_2 = 1.255$, $1.259$, respectively.  The latter
minimum corresponds to a curvature radius $r_0 = \sqrt{12/0.045}
\approx 16$, which is not unreasonable compared to the previous 13.

We conclude that for $\kappa_2 = 1.255 - 1.259$ the spacetimes are on
the average near $S^4$ in the distance regime $6 \leq r\leq 24$.

\section{Binding in dynamical triangulation}

On each dynamical triangulation configuration we calculated the
propagator of the scalar field
\begin{equation}
  G_{xy} = (-\square + m_0^2)^{-1}_{xy},
\end{equation}
using the algebraic multigrid routine AMG1R5, where $x$ is an
arbitrary origin.  The discrete Laplacian is defined as
\begin{equation}
  (\square)_{xy} =
  \left\{\begin{array}{cc}
    1 & \mbox{if $x$ and $y$ are nearest neighbours} , \\
   -5 & \mbox{if $x = y$} , \\ 
    0 & \mbox{otherwise} .
  \end{array}\right.
\end{equation}
The $5$ in the second line arises as the coordination number of a
four-simplex, i.e.\ a four-simplex has five neighbours.

We can then calculate $G(r)$ and $G^{(2)}(r)$ by averaging $G_{xy}$
respectively its square over all points $y$ at distance $r$ from the
origin $x$, and then over origins and configurations,
\begin{eqnarray}
G(r) &=& \left\langle  
                      \frac{\sum_y G_{xy} \delta_{d_{xy},r} }
                           {\sum_y        \delta_{d_{xy},r} }
         \right\rangle_g,
\label{simpleG}\\
G^{(2)}(r) &=&
         \left\langle  
                      \frac{\sum_y G_{xy}^2 \delta_{d_{xy},r} }
                           {\sum_y          \delta_{d_{xy},r} }
         \right\rangle_g
\label{simpleG2}
\end{eqnarray}
(we have not indicated the average over origins $x$).  Notice that
(\ref{simpleG}) corresponds to (\ref{scalprop}) in the continuum.

To improve the calculation of the binding energy, we can try to use
what are called ``smeared sources''.  The use of these smeared sources
can improve the data by increasing the contribution of the ground
state and decreasing the contribution of the excited states.  Instead
of using (\ref{simpleG2}), we can calculate $G(r)$ by averaging
$G_{xy}$ over all points at distance $r$ from the origin $x$, and only
after taking this average, square it for the calculation of
$G^{(2)}(r)$:
\begin{equation}
G^{(2)}(r) = \left\langle
             \left(\frac{\sum_y G_{xy} \delta_{d_{xy},r} }
                        {\sum_y        \delta_{d_{xy},r} }
             \right)^2
             \right\rangle_g .
\label{improG2}
\end{equation}
This corresponds to taking the propagator from a source that is not a
single point, but a complete shell around the origin.  Such a source
may have a bigger overlap with the ground state wave function and a
smaller one with the excited wave functions.  For a discussion of the
use of these smeared sources in QCD, see e.g.\ \cite{Gu90}.

\begin{figure}[t]
  \dataplot{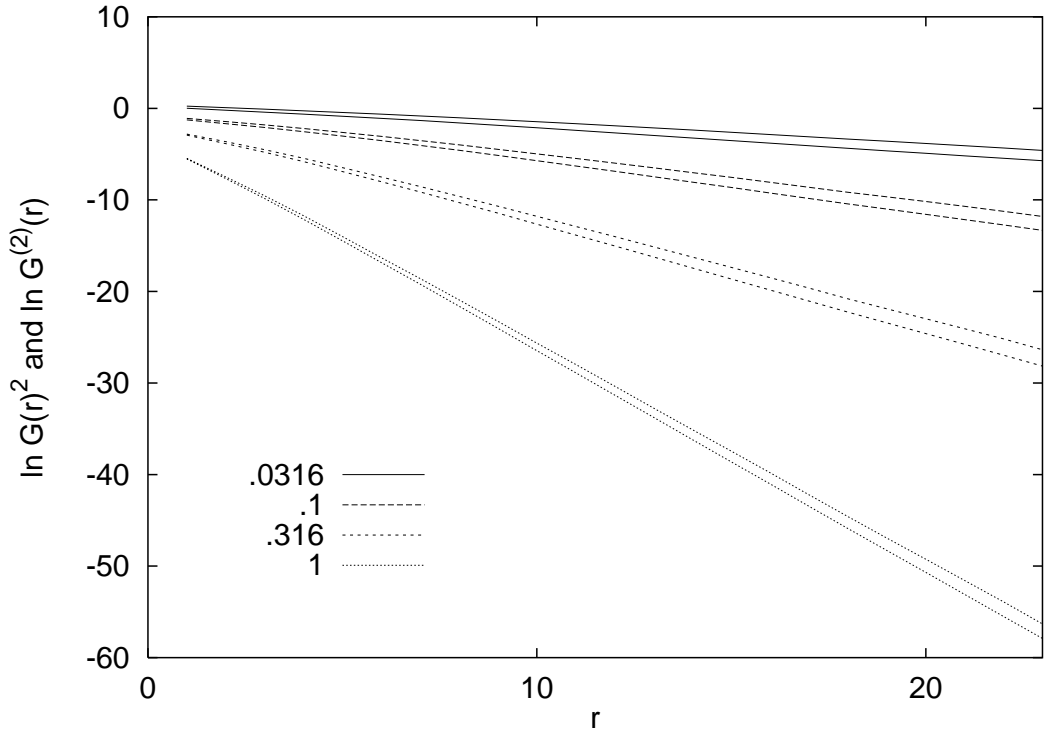}
  \caption{The two-particle propagator and the square of the one
    particle propagator versus the geodesic distance for four
    different bare masses $m_0$.  The vertical scale is logarithmic.
    $\kappa_2 = 1.255$, which is very close to the transition.}
  \label{4dprop}
\end{figure}

Because the average of the square of a fluctuating quantity is always
greater than the square of its average, it is obvious that $G^{(2)}(r)
> G(r)$.  This does not yet imply anything about the way they fall
off.  In particular it is not guaranteed that the energy of the
two-particle state is less than twice the energy of the one particle
states.

In figure \ref{4dprop} we see the results for four different bare
masses. Each pair of lines corresponds to one bare mass.  In each pair
the upper line is $G^{(2)}(r)$ (using (\ref{simpleG2})) and the lower
line is $G(r)^2$ of eq.~(\ref{simpleG}).  We used $144$ configurations
recorded every $5000$ sweeps (1 sweep = $N$ accepted moves).  For each
of the masses we used $120$ origins per configuration.  The coupling
constant $\kappa_2 = 1.255$, which is the lower of the two $\kappa_2$
values used in the $N'(r)$ figure~\ref{s4fig}.

There is clearly a difference in slope between the lines in each pair.
This shows that the energy of the two particle state is less than two
times the mass of a single particle and consequently that there is a
positive binding energy between the particles.

Using this data we can measure the renormalised mass $m$, by assuming
a long distance behaviour of $r^{\alpha}\exp(-mr)$.  The results are
(in parenthesis is the value of $\kappa_2$)
\begin{equation}
  \begin{array}{c|c|c}
    m_0 & m(1.255) & m(1.259) \\ \hline
    0.0316 & 0.14 & 0.12  \\
    0.1 & 0.29 & 0.27  \\
    0.316 & 0.60 & 0.58  \\
    1 & 1.21 & 1.20 
  \end{array}
\label{mm0}
\end{equation}
It was argued in \cite{AgMi92b} that the physical mass should vanish
at zero bare mass and that therefore the renormalisation would be only
multiplicative.  Our data seem to show that the relation is more
complicated.  Increasing $m_0$ by a factor of $\sqrt{10} \approx 3.16$
increases $m$ by a factor of about $2.1$.

Comparing figure~\ref{curvature} with figure~\ref{4dprop}, we see that
the long distance behaviour is indeed similar, being an exponential.
The finite size effect of the $S^4$-like curvature is apparently
negligible, except perhaps for the smallest mass.  The short distance
behaviour is quite different.  The propagators in figure~\ref{4dprop}
curve downward towards the origin, while the free propagators shown in
figure~\ref{4dprop} curve upwards due to the $1/r^2$ behaviour.  The
curving downward is unusual, because a propagator is interpreted as a
sum of decaying exponentials corresponding to the ground state and
various excited states.  A closer look shows that the downward
curvature occurs for distances smaller than about $r=5$, which is
roughly the end of the `planckian regime' mentioned in the previous
section.  As we have seen in figure~\ref{curvature}, larger positive
curvature means that the propagator decreases more slowly.  Therefore,
such a planckian regime may cause the propagator to decrease less at
smaller distances where the effective curvature is large than at the
longer distances where the effective curvature is small.

Using these data, we can now estimate the binding energy of the
particles.  From (\ref{gofd}) and (\ref{g2ofd}) we have
\begin{eqnarray}
E_b(r) & \equiv & r^{-1} \ln \frac{G^{(2)}(r)}{G(r)^2}\\
       & \to    & E_b = 2m-M, \qquad r \to \infty .
\end{eqnarray}
As we cannot use infinite distances, we will consider the effective
binding energy $E_b(r)$ for finite $r$ and look whether this
expression becomes constant.

\begin{figure}[t]
  \dataplot{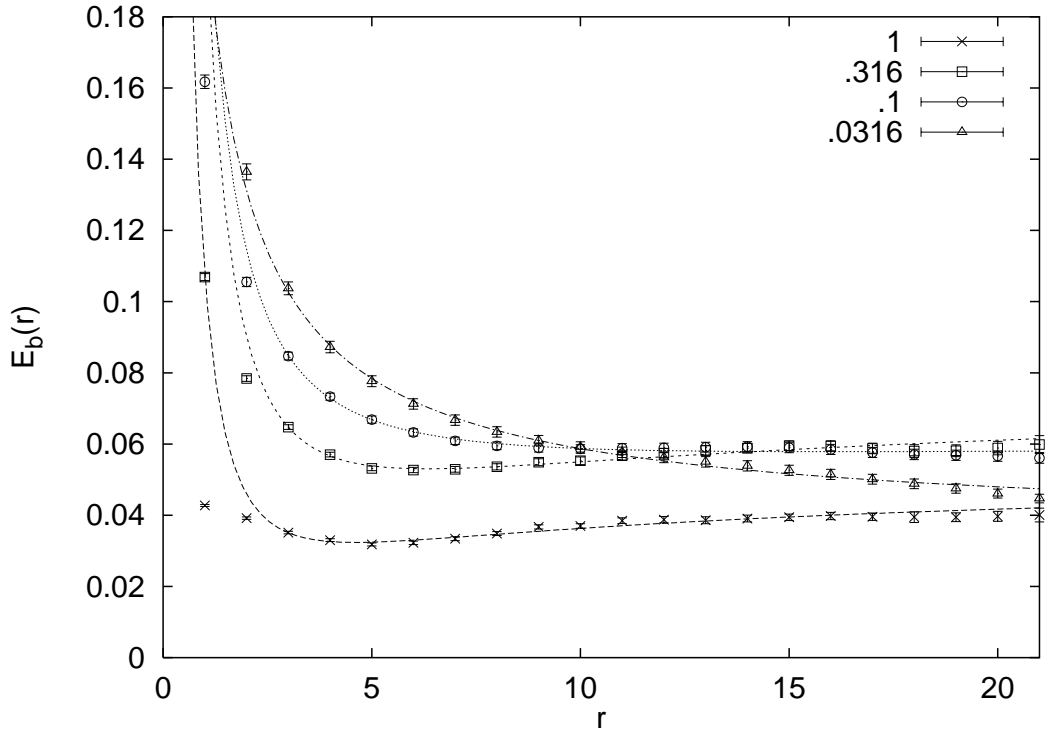}
  \caption{The effective binding energy $E_b$ as a function of the
    geodesic distance for four different bare masses at $\kappa_2 =
    1.255$. The lines are fits tot the data with the form 
  \protect (\ref{Ebfit}).}
  \label{be255}
\end{figure}

Figure~\ref{be255} shows this quantity as a function of the geodesic
distance, using the smeared estimator (\ref{improG2}) for
$G^{(2)}(r)$. The four curves again correspond to the four different
bare masses in figure~\ref{4dprop}.  To avoid the correlations between
origins on the same configuration and between points at the same
distance of such an origin influencing the error bars, we first
averaged all the measurements of each configuration and used a
jackknife method on these averages to calculate the error bars.

It is clear that the binding energy goes to a non-zero value, with the
exception of the lowest mass, where the effective binding energy does
not seem to converge within the limited distance range.  We chose not
to display distances larger than the position of the maximum of
$N'(r)$ in figure~\ref{s4fig}.

Unfortunately, the correlation between the mass and the binding energy
does not appear to be strictly positive.  The lowest binding energy
belongs to the highest mass.  We defer a more elaborate discussion to
the next section.

\begin{figure}[t]
  \dataplot{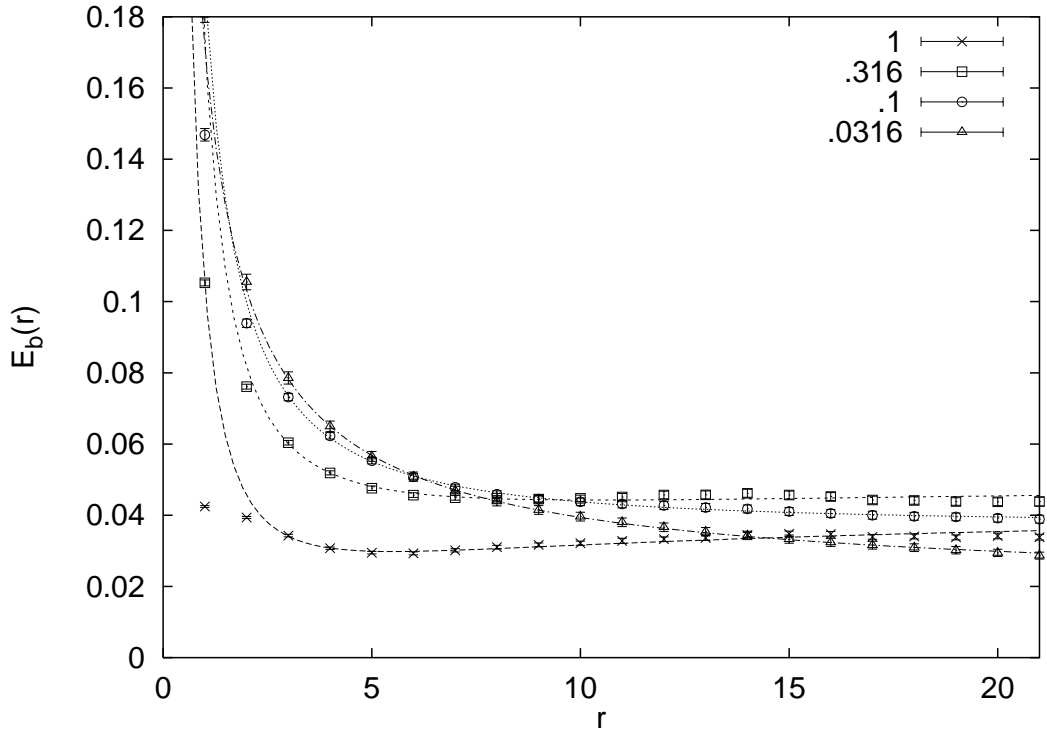}
  \caption{Like figure \protect\ref{be255}, but with $\kappa_2 =
    1.259$.}
  \label{be259}
\end{figure}

Figure~\ref{be259} shows the corresponding data for $\kappa_2 = 1.259$
which is the higher of the two $\kappa_2$ values used in
figure~\ref{s4fig} for $N'(r)$.  In this case we used $200$
configurations, again with $120$ origins per configuration.  The data
look very similar.  Here the ordering of binding energies follows more
clearly that of the constituent masses, except for the largest mass
$m_0 = 1$.

\begin{figure}[t]
  \dataplot{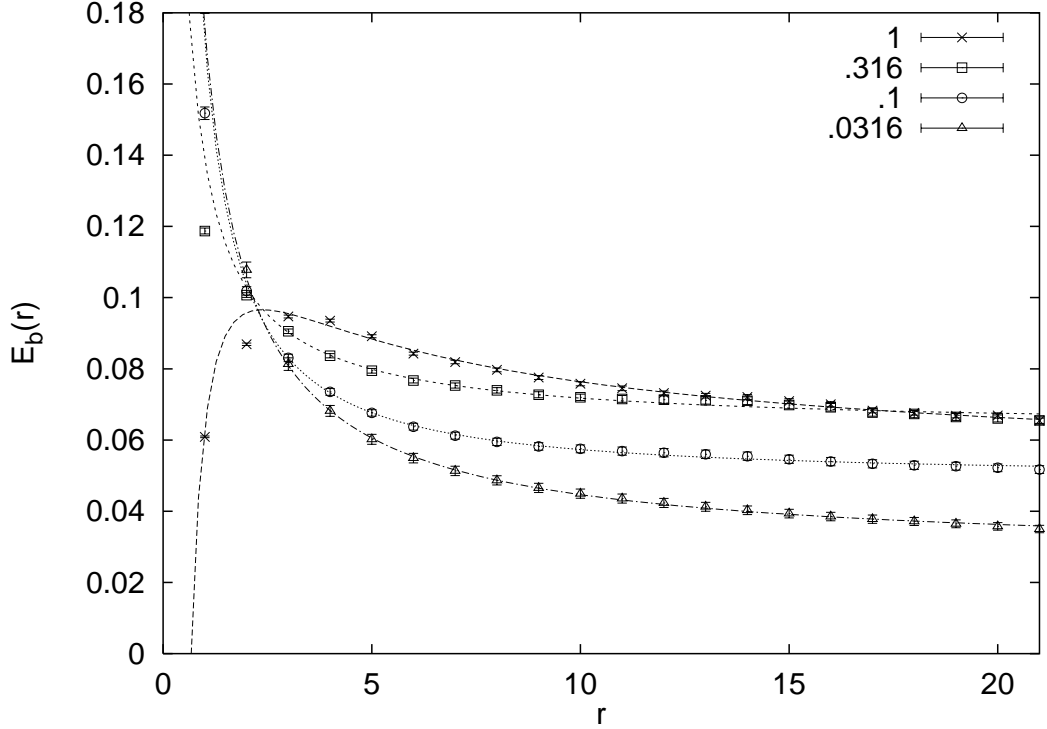}
  \caption{Like figure \protect\ref{be259}, but with unsmeared
    sources.}
  \label{be259unsm}
\end{figure}

To see the effect of using smeared sources we show in
figure~\ref{be259unsm} the effective binding energy using the
unsmeared estimator (\ref{simpleG2}) for $\kappa_2=1.259$, the
analogue of figure~\ref{be259}.  The use of smeared sources
(figure~\ref{be259} compared to figure~\ref{be259unsm}) does indeed
improve the convergence to a definite value and the effective binding
energies are generally smaller.  In particular the curve for a bare
mass of $1$, which keeps going down in figure~\ref{be255} becomes more
horizontal.  Only the smallest mass is an exception, where neither
case has converged yet in the distance range shown.

To determine the binding energies we have fitted the (smeared) 
effective $E_b(r)$ to the form
\begin{equation}
E_b(r) = E_b + (c+\gamma\ln r)\, r^{-1},
\label{Ebfit}
\end{equation}
which corresponds to the asymptotic forms (\ref{gofd}) and (\ref{g2ofd}) with 
$c = \ln (Z^2 / Z^{(2)})$ and $\gamma = 2\alpha - \beta$. In general one would expect
$c$ to be positive since $\phi^2$ will be relatively less effective in creating the 
two-particle bound state than $\phi$ the single particle state. 
If the single and two-particle 
propagators behave sufficiently like a scalar propagator in flat space we would have 
$\alpha \approx \beta < 0$ , hence also $\gamma < 0$. The results fo the
fits in the region $3 \leq r \leq 20$ (recall $r_m \approx 21 - 22$) are shown in the
following table. 
\begin{equation}
  \begin{array}{l|l|l|l|c|l|l}
m_0 & E_b(1.255) & c(1.255) & \gamma(1.255) & E_b(1.259) & c(1.259) & \gamma(1.259) \\ 
\hline
1      & 0.054(1) & 0.053(5) & -0.100(7) & 0.045(1) & 0.059(4) & -0.084(6) \\
0.316  & 0.078(2) & 0.134(9) & -0.16(1)  & 0.053(1)  & 0.19(7)  & -0.089(9)  \\
0.1    & 0.064(2) & 0.17(1)  & -0.09(1)  & 0.038(2)  & 0.15(1)  & -0.04(1)  \\
0.0316 & 0.035(2) & 0.17(2)  & +0.03(2)  & 0.019(2)  & 0.15(2)  & +0.02(2)
  \end{array}
\label{Eb}
\end{equation}
The signs of $c$ and $\gamma$ appear to follow the above expectations, 
except for $\gamma$ for the smallest bare mass.  
The corresponding effective binding energy is still falling
for the distances shown and has to be viewed with caution. 
We cannot improve on
this because we do not trust distances much beyond the maximum
of $N'(r)$ since these run into its asymmetric tail.

\section{Discussion}

The spacetimes produced by
the dynamical triangulation model have semiclassical features 
near the phase transition.
For the binding energy computations we have chosen two values of $\kappa_2$ 
such that the volume at distance $r$, $N'(r)$, behaves like
that of a four-sphere, for not too large distances $r$ outside a
`planckian regime'. 

Let us briefly recall here
our scenario for reducing the lattice distance \cite{BaSm95a}:
increasing $N$ we have to tune $\kappa_2$ such that we stay on the same
scaling function $\rho = r_m N'(r)/N$. This can be tried for different
starting points $(\kappa_2,N)$, corresponding to different shapes of $\rho$ as
a function of $r/r_m$. 
The scaling of $\rho$ has to be carefully re-examined, especially now the 
phase transition appears to be first order.

The data for a nonzero binding energy are quite convincing and encouraging.
For given bare mass $m_0$, the resulting renormalized masses $m$ are closer
to $m_0$ and the binding energies turn out to be smaller for the larger 
$\kappa_2$ value (1.259), than for the smaller value (1.255). 
This suggests that the renormalized Newton constant $G$
is smaller for the larger $\kappa_2$ value,
assuming of course $G$ exists and is positive for these $\kappa_2$ values.
Since $G_0 \propto 1/\kappa_2$ and one would indeed expect $G$ to
decrease when $\kappa_2$ increases.

We would like to be able to extract 
$G$ from the data, for example according to the nonrelativistic
formula
\begin{equation}
  E_b \equiv 2m - M = \frac{1}{4} G^2 m^5 .
  \label{hydrogenlike}
\end{equation}
This formula is just the familiar energy $\alpha^2 m_{\rm red}/2$ of
the hydrogen atom in the ground state, but with the gravitational
parameters substituted as $\alpha \to G m^2$ and the reduced mass
$m_{\rm red} \to m/2$.  Because the formula (\ref{hydrogenlike}) is
nonrelativistic it may not suffice to fit the data. To get a
rudimentary feeling for corrections to (\ref{hydrogenlike}) we
consider the hamiltonian
\begin{equation}
H=2\sqrt{m^2 + p^2} - Gm^2/r.
\end{equation}
Replacing $p\rightarrow 1/r$ and minimising the energy leads to 
\begin{equation}
  E_b = 2m-2m\sqrt{1-G^2 m^4/4},
  \label{berel}
\end{equation}
which suggests that $Gm^2 = 2$ has to be considered `large'.

Unfortunately, the data in figure \ref{be255} or \ref{be259} show no
sign of the $m^5$ behaviour of (\ref{hydrogenlike}), and neither is
(\ref{berel}) of any help.  Even the largest constituent mass $m
\approx 1.2$, which is evidently so large in lattice units that only
qualitative conclusions may be drawn from it, leads to a small binding
energy.  For the lighter constituent masses the binding energy is only
modestly dependent on $m$.

Perhaps the behaviour for the lightest constituent mass hints at a
possible interpretation. If we make
the bold assumption that the nonrelativistic formula
(\ref{hydrogenlike}) starts making sense for the lightest constituent
masses, 
(\ref{mm0},\ref{Eb}) lead to a renormalised Planck length
$\sqrt{G} = 7.1$ and 7.4 respectively for $\kappa_2 = 1.255$ and 1.259. 
This appears to contradict the suggestion above that $G$ is smaller for the 
larger $\kappa_2$ value, but the uncertainties in $m$ are greatly magnified 
by taking its fifth power.
The values for $\sqrt{G}$ are reasonable and they are 
furthermore similar to the (somewhat vague) distance scale the
planckian regime ends.  On the other hand, the size of a
nonrelativistic bound state is of order of $1/m$, so if indeed
$\sqrt{G} \approx 7$, we should perhaps not be surprised to find odd
behaviour for bound state sizes of the order of the Planck length, or
constituent masses greater than the Planck mass $1/\sqrt{G} \approx
0.14$.

Clearly, we are having a problem of separating scales: we would like
$\sqrt{G} \ll m^{-1} \ll r_m$, where $r_m = r_0 \pi/2$ is a measure of
the size of our $S^4$-`universe' (the distance where the volume
$N'(r)$ is maximal). In our situation, at best, $\sqrt{G} \approx
m^{-1}$ and $r_m \approx 21 \approx 3 m^{-1}$.  The size of our
universe is only three times the Planck length.  It is essential for
a physical 
interpretation that as the lattices get bigger, the planckian
regime and $\sqrt{G}$ shrink in units of $r_m$.  We found some evidence for this in
our scaling analysis \cite{BaSm95a} where we ventured a scenario in
which $\sqrt{G}/r_m \to 0$ because of triviality.  This approach to
zero might then be only logarithmic, which would make the problem of
scale separation severe from the computational point of view.  A
careful study is needed to clarify these issues.
If the separation of scales would not materialize we could still 
study the continuum behaviour of the complete function
$G^{(2)}(r)$ as a welcome addition to the $N'(r)$ used so far.

\section*{Acknowledgements}

We thank Piotr Bia{\l}as for useful discussions.  The numerical
simulations were carried out on the IBM SP1 at SARA and the Parsytec
PowerXplorer at IC3A.  This work is supported in part by FOM.

\end{document}